\documentclass{sigchi}
\CopyrightYear{2020}
\setcopyright{acmlicensed}
\doi{http://dx.doi.org/10.1145/3313831.3376643}
\isbn{978-1-4503-6708-0/20/04}
\conferenceinfo{CHI '20,}{April 25--30, 2020, Honolulu, HI, USA}
\acmPrice{\$15.00}

\usepackage{balance}       %
\usepackage{graphics}      %
\usepackage[T1]{fontenc}   %
\usepackage{txfonts}
\usepackage{mathptmx}
\usepackage[pdflang={en-US},pdftex]{hyperref}
\usepackage{color}
\usepackage{booktabs}
\usepackage{textcomp}
\usepackage{microtype}        %
\usepackage{ccicons}          %
\usepackage[utf8]{inputenc} %
\usepackage{enumitem}

\usepackage{subfig}
\usepackage[font={small,bf},skip=2pt]{caption}
\usepackage{multirow} %
\usepackage[table]{xcolor} %

\def\plaintitle{Physiologically Driven Storytelling:\\ Concept and Software Tool}
\def\plainauthor{Jeremy Frey, Gilad Ostrin, May Grabli, Jessica R. Cauchard}
\def\plainkeywords{Affective Computing; Physiology; Interactive Fiction; Storytelling; Taxonomy; Physiological Computing}

\makeatletter
\def\url@leostyle{%
  \@ifundefined{selectfont}{
    \def\UrlFont{\sf}
  }{
    \def\UrlFont{\small\bf\ttfamily}
  }}
\makeatother
\def\pprw{8.5in}
\def\pprh{11in}

\definecolor{linkColor}{RGB}{6,125,233}
\hypersetup{%
  pdftitle={\plaintitle},
  pdfauthor={\plainauthor},
  pdfkeywords={\plainkeywords},
  pdfdisplaydoctitle=true, %
  bookmarksnumbered,
  pdfstartview={FitH},
  colorlinks,
  citecolor=black,
  filecolor=black,
  linkcolor=black,
  urlcolor=linkColor,
  breaklinks=true,
  hypertexnames=false
}

\teaser{ 
\centering 
\includegraphics[width=1\textwidth]{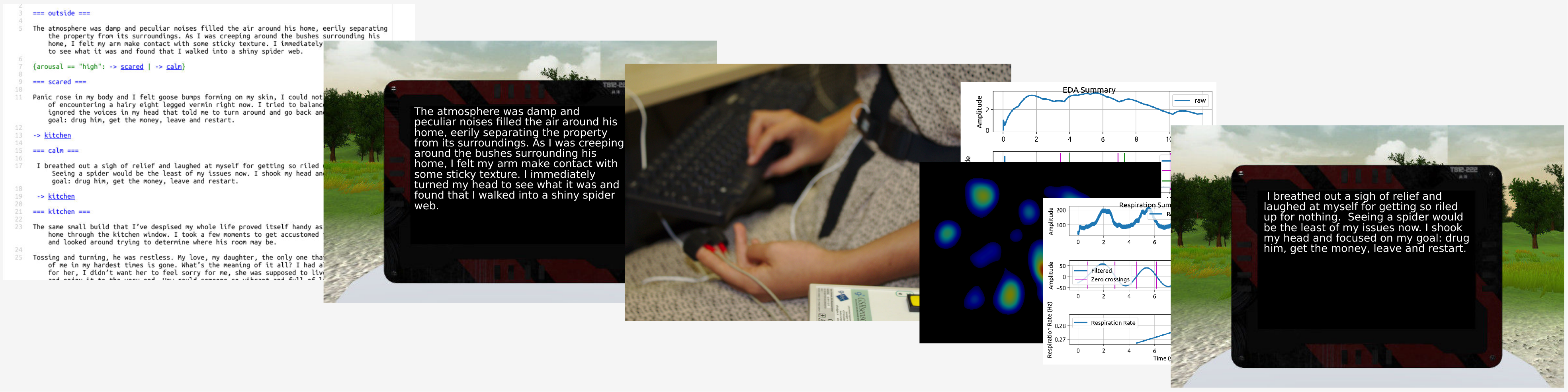} 
\caption{A story is rendered with PIF while physiological signals are monitored and analyzed in the background to comprehend the reader's experience. The detection of specific states can be used to automatically change the story and augment reading experience as the interactive story unfolds.} 
\label{fig:teaser} 
}

\begin{document}

\title{\plaintitle}

\numberofauthors{1}
\author{%
  \alignauthor{J\'er\'emy Frey$^{1,3}$, Gilad Ostrin$^{2,3}$, May Grabli$^{3}$, and Jessica R. Cauchard$^{2,3}$\\
    \affaddr{$^{1}$Ullo, La Rochelle, France}\\
    \affaddr{$^{2}$Magic Lab, Ben Gurion University of the Negev, Be'er Sheva, Israel}\\
    \affaddr{$^{3}$Interdisciplinary Center (IDC) Herzliya, Herzliya, Israel}\\
    \email {jfrey@ullo.fr, gilad.ostrin@gmail.com, maygrabli@gmail.com, jcauchard@acm.org}\\}
    }

\maketitle

\begin{abstract}

We put forth Physiologically Driven Storytelling, a new approach to interactive storytelling where narratives adaptively unfold based on the reader's physiological state. 

We first describe a taxonomy framing how physiological signals can be used to drive interactive systems both as input and output. We then propose applications to interactive storytelling and describe the implementation of a software tool to create Physiological Interactive Fiction (PIF). The results of an online study (N=140) provided guidelines towards augmenting the reading experience. PIF was then evaluated in a lab study (N=14) to determine how physiological signals can be used to infer a reader's state. Our results show that breathing, electrodermal activity, and eye tracking can help differentiate positive from negative tones, and monotonous from exciting events. 

This work demonstrates how PIF can support storytelling in creating engaging content and experience tailored to the reader. Moreover, it opens the space to future physiologically driven systems within broader application areas.

\end{abstract}




\section{Introduction}
Stories have the power to open people's minds. Yet, our life experience and preconceptions affect how we relate to them. Moreover, the mood we are in can change which book we decide to read. Inversely, the narrative itself can alter how we feel and perceive the world \cite{Mar2017}. We propose to combine modern storytelling techniques and physiological data to broaden the perspective of both readers and writers. %
We suggest that by taking into account how a reader feels and how a story affects them, it is possible to tailor the narrative on the fly. 

Thanks to the recent increase and availability of physiological sensors in mobile and wearable consumer products, we find ourselves at a unique time to include physiology into wider applications. 
We propose that physiology can be used as input or output in interactive narration. For example, as \textit{explicit input}, the reader would be acting on their physiology to achieve a particular goal, while as \textit{implicit input}, the system would determine which emotions elicit specific events. For example, a storyline could gain intensity as the reader gets scared. 

This work describes the design, development, and evaluation of a software tool for physiologically driven storytelling named PIF: Physiological Interactive Fiction (Figure \ref{fig:teaser}). PIF embeds, for the first time, a reader's state detection and real-time text adaptation. By automatically changing the story depending on the reader's state, PIF enables writers to create adaptive narratives to improve the reading experience. We suggest that adapting a character's reaction to the reader's emotion could increase identification and immersion, which in turn could support the reader in connecting and empathizing with the characters and situations they read about \cite{Cohen2015,Graaf2012}. While the concept of physiologically driven interactive fiction is potentially applicable to all media, this work focuses on text-based narratives. One major advantage of text-based stories, over graphics, is that readers can use their imagination to fill in the gaps. Indeed, text is a powerful medium, which can easily be modified, compared to video footage or graphics which are difficult to modify once created. Moreover, while a single writer can embed a whole world in a book, creating movies or video games require large team effort.  %

PIF combines two visionary perspectives on the use of digital tools: \textbf{interactive narratives} as a way to relate to fictive or real-life events \cite{Murray1997}; and \textbf{physiological measurements} as a way to assess and affect people \cite{Picard1995}. 
PIF is built upon an established markup language and made available as an open-source tool to provide writers with a novel way to express themselves. Combined with machine learning, PIF allows for real-time processing of physiological signals. 

This paper first describes the related work and presents a taxonomy of the use of physiological signals in interactive systems. It then presents the design, implementation, and evaluation of the PIF software tool. As part of the software tool design, we investigated a set of physiological signals that allow detecting which specific cognitive or affective states can be used as implicit input to automatically adapt a story. An online study (N=140) was performed to provide guidelines on how to augment the reading experience. Results showed how a character's reactions to negative stimuli is a predictor of narrative engagement and perceived similarity (i.e., the degree to which the reader feels similarity with the character). To demonstrate how physiology can account for a reader's inner state, we then performed a laboratory study (N=14). Using machine learning on physiological signals, such as breathing and eye tracking%
, we find that classifying stories between positive and negative tones, and between boring and exciting narratives, yielded promising performances. This work's contributions to the field are as follows:

\begin{itemize} %
\item Taxonomy of physiologically driven interactive systems.
\item PIF: Physiological Interactive Fiction software tool for interactive narratives using physiological signals as both input and output. %
\item A user study (N=140) which determines features that can be used to adapt a story to the reader.
\item A lab study (N=14) which classifies a reader's physiological reactions to stories through the measurement of breathing, electrodermal activity, and eye tracking. 
\end{itemize}

\section{Related Work}

This section presents prior work on interactive storytelling, including a discussion on input techniques, and existing research and commercial authoring platforms to create such content.

\subsection{Interactive Fiction}

Interactive Fiction (IF) is a literature genre in which the storyline is not predetermined but evolves based on the reader's input. Emerging in the late 70s, IF encompasses \textit{Choose-Your-Own-Adventure} stories, where readers navigate branches to advance a narrative, and games where users interact with the textual environment~\cite{Montfort2003}. IF stories have been designed using all forms of media, from simple text to complex animated movies (e.g., Dragon's Lair \cite{DragonsLair}). However, creating complex content is not only time-consuming, it is also difficult to modify after the creation stage.
Peng et al. \cite{Peng2018} denote how a change, as subtle as the trajectory of a 3D avatar, can induce complex computations and impede the variety of adaptations. Seeking to provide a tool for self-reflection, the authors detail that such computational complexity prevented them from achieving their goal, here associating stress with the appearance of a mountain in front of the main character to symbolize an obstacle. However, in text-based narratives, readers can picture various situations with a simple change in well-chosen words, without the need for complex modifications. As such, we decided to use text as a medium for this work. 

Various input mechanisms have been proposed, over the years, to let readers explore IF storylines. Computer games, such as ``Adventure'' \cite{Adventure} and Infocom's Zork \cite{Zork}, had players interact using natural language by typing text commands on the keyboard (e.g., ``GO NORTH''). Nowadays, most IF games rely on explicit choice selection through mouse clicks. Readers select the desired story branch on the screen, as with the visual novel engine Ren'Py\cite{siteRenpy}, which presents a sequence of images alongside text, or Inkle's ``Sorcery!'' game \cite{SorceryGame}, adapted from the eponymous Choose-Your-Own-Adventure book \cite{SorceryBook}.

\begin{table*}[ht!]
\centering
\begin{tabular}{c|c|c|c}
 Input & Medium & Authoring Tool & Notes \\
\hline

\hline
 \cellcolor{green!10} & \cellcolor{green!10}text &  \cellcolor{green!10} markup language &  \cellcolor{green!10} PIF: open source research project \\
 & text &  \textit{N.A.} & \cite{Ismail2011,Barral2017}: research project, text annotation (no adaptation) \\
 & multimedia &  \textit{N.A.} & \cite{Chanel2011}: research project, adapts game difficulty \\
 & 3D & \textit{N.A.} & \cite{Gilroy2012,Cavazza2016}: research project, detects empathy to adapt content \\
 \multirow{-5}{*}{ physiology } & 3D & \textit{N.A.} & \cite{Lobel2016}: commercial game, adapts horror level \\

\hline
 &  text & \textit{N.A.} & \cite{Murray1999}: research project, adapts learning exercises difficulty \\
 \multirow{-2}{*}{ user behavior } & 3D & \textit{N.A.} & \cite{Yannakakis2013}: commercial game, adapts difficulty \\

\hline
 survey & 3D & \textit{N.A.} & \cite{Peng2018}: research project, how to engage viewers \\

\hline
 & text & visual programming & Twine: open source community project, publishes to web \\
 & text &  markup language & Ink: open source project, released by Inkle studio  \\
 \multirow{-3}{*}{ choice based } & multimedia & programming & Ren'Py: open source community project, visual novels \\

\hline
 &  text &  markup language & Inform: open source community project, compiles to z-code  \\
 \multirow{-2}{*}{ natural language } & text & \textit{N.A.} & Infocom: commercial games, created z-code interpreters \\

\end{tabular}

\caption{Classification of existing commercial and academic IF tools by input type, medium, and availability of an authoring tool. \emph{N.A.}: Not Available.}

\label{tab:tax}
\end{table*}

Recently, alternative inputs methods have been proposed to enable immersive narration and enhance the reading experience. Peng et al. \cite{Peng2018} proposed 3D animated short stories tailored to viewers, using self-report and behavioral data, to induce engagement. Their work represents a step towards natural and seamless interaction, which could also be performed using physiological measurements in combination with machine learning approaches, as to estimate a variety of states, such as emotions, attention level, and cognitive workload (see \cite{Cowley2016} for a review). Such approach was explored by Cavazza and Charles \cite{Cavazza2016} who used brain recordings to detect empathy and alter a narrative, building upon a framework aimed at adapting 3D narrative based on physiology \cite{Gilroy2012}. In gaming, monitoring heart-rate has been used to alter the gameplay, such as to encourage the user to remain calm while playing a horror game \cite{Lobel2016}. Brain recordings have been used to adapt game difficulty and maintain a state of flow, a mental state in which a person is highly engaged in a task, losing the perception of time \cite{Chanel2011}. Brain recordings and facial expressions were also used in combination, using textual stimuli, to investigate whether emotions can be annotated \cite{Ismail2011} and to detect the relevance of a piece of information during a reading session \cite{Barral2017}.

In these prior works, physiological signals were recorded for offline analysis at a later stage. In contrast, this present work aims at using physiology as a novel real-time input strategy. %

\subsection{Interactive Fiction Authoring tools}

The possibility to openly create, share, and write interactive fiction combined with the access to open-source authoring tools are the main reasons why the IF community strives \cite{Montfort2012}. Providing the right environment to content creators is a prerequisite to the adoption and dissemination of this literature genre. In this subsection, we review existing authoring tools and examine how PIF can leverage existing technologies to incorporate physiology to current writing tools. Inform \cite{siteInform}, one of the first authoring tools, eased content creation by proposing a language closer to markup than to programming. It facilitated the diffusion of interactive stories by compiling to z-machine instructions that could run on the interpreters developed by Infocom, for all major platforms at the time. More recently, Twine \cite{siteTwinery} proposed a web-based graphical programming interface to write interactive fiction. Moreover, research in interactive fiction and natural language processing resulted in projects where a whole text could evolve based on the user's input, as in \cite{Montfort2011}. Yet, such complex environments are difficult to conceive due to the important number of rules that must be programmed. As such, in this first step towards implicit adaptation, we focus on branching narratives that are easier to write and accessible to most authors.

Table \ref{tab:tax} summarizes the prior IF tools divided across three dimensions: 1. type of input 2. medium and 3. the type of authoring platform. The table highlights a gap in the literature, as no authoring tool currently exists for the creation of a narrative that can automatically adapt to the reader. This motivated our work and the design of PIF as a tool which relies on a simple markup language and which uses physiology as an alternative input to explicit interactions from the reader, extending on earlier proofs of concept \cite{Frey2016f,10.1145/3282894.3289740}. To favour adoption and dissemination of physiologically driven narratives, PIF is based on Inkle's Ink \cite{siteInkle}, which can be ported to web or phone environments. Ink can be bridged with modern development environments such as Unity for mobile, web, and virtual environments. 
With an active community behind it, Ink was a perfect candidate for our new engine PIF. 

In the next section, we describe a taxonomy of the use of physiological signals in interactive systems which helped inform the design of PIF and explore all potential use cases of physiologically driven storytelling.

\section{Taxonomy: The Tale of Four Usages}

Physiological signals have been used in a wide range of applications to monitor, as well as supplement, motor control in disciplines ranging from medicine to sports and entertainment. To understand all potential usages of physiological signals in interactive systems, we propose a taxonomy which defines the role of the user with regards to the interactive system, based on how physiology is being used (i.e., as input or output). Prior work advanced classifications pertaining to physiological computing, %
such as the distinction between: active, reactive, and passive Brain-Computer Interfaces (BCI) \cite{nicolas2012brain}. This taxonomy differs from prior work as it considers, not only physiology as input, but also as output from the interactive system, showing how both input and output can affect the user.

\begin{figure}[t]
\centering
\includegraphics[width=0.9\hsize]{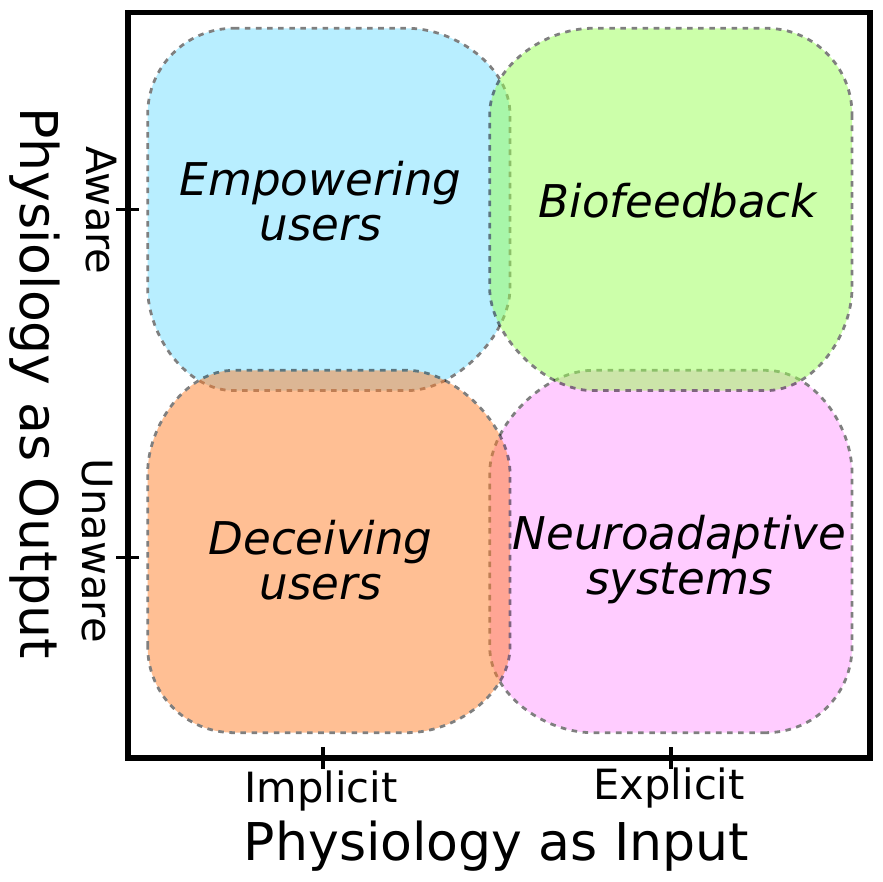}
\caption{Taxonomy of physiologically driven interactive systems, along two axes: Physiology as Input (implicit or explicit) and Physiology as Output (with users aware or unaware). The combination of these two dimensions define four usage scenarios, all supported by PIF.}
\label{fig:taxonomy}
\end{figure}

The taxonomy (Figure \ref{fig:taxonomy}) comprises two axes: \textbf{Physiology as Input}, meaning that physiological signals are read by the system, and \textbf{Physiology as Output} where the system alters the physiology of the user. Each axis in divided into two categories such that: Input can be either \textit{implicit}, when the adaptation occurs in the background or \textit{explicit} if users have to consciously change their physiology. Output is then divided depending on whether users are \textit{aware} that the system might be affecting them in return or \textit{unaware} of it. We find that all physiologically driven systems fall under one of the four quadrants. The subsections below detail scenarios for each use-case and help inform the design of PIF.

\subsubsection{Input: Explicit, Output: Aware - Biofeedback}

In this quadrant, the input is explicit and the user knows that their physiology is being monitored and affected. This corresponds to applications where physiology is used to help people regulate their state. For example, \textbf{biofeedack} can be used to observe and act upon specific brain activities and alleviate stress, as in \emph{neurofeedback}. Yucha \& Montgomery \cite{Yucha2008} presented a review of such use in medical settings. In HCI, biofeedback enables alternative communication channels, such as sending and receiving real-time breathing patterns with a relative \cite{Frey2018}. In entertainment, it can be used to alter gameplay \cite{Nacke2011a} or seek emotion regulation \cite{Lobel2016}. In PIF, readers could voluntary change their physiological state to navigate through the story. For example, a reader could perform breathing exercises so the character could enter a monastery.

\subsubsection{Input: Explicit, Output: Unaware - Neuroadaptive systems}

This category depicts situations where the user knows that the system is measuring their physiology, but they are not aware that the system reacts to this information and adapts accordingly. In an HCI context, this category encompasses \textbf{neuroadaptive interfaces} \cite{Krol2017} (also known as passive BCI). Such systems can automatically evolve to better suit users, such as when matching a task to the user's workload \cite{Mehta2013}. They can also be used to improve safety, such as in aviation, where a pilot's state can be monitored during flights \cite{Dehais2019}. In PIF, the system could monitor the reader's state to match the reading difficulty. This could be used for textbooks where the lesson's difficulty would adapt to the student's needs, alleviating psychological stress. Indeed, prior work hypothesized that physiologically driven tutoring systems might better adapt content and ease the workload \cite{Gilbert2011}.

\subsubsection{Input Implicit, Output Unaware - Deceiving users}

Applications in this category occur covertly, altering the user's states without them being aware of it. This type of applications are similar to advertisement companies tracking people's behaviors to trigger compulsory purchase; but this time, the input would be the user's physiological state. While it might be possible to covertly manipulate a person's state, this poses ethical concerns around \textbf{deceiving users} \cite{Ienca2017}, making this quadrant the most controversial use of physiologically driven interactive systems. These types of usage would correspond to influencing a person without their awareness and consent, which is outside of the ethical scope of our research project.

\subsubsection{Input: Implicit, Output: Aware - Empowering users}

Here the user is aware that the system may affect them, without realizing how the input is actually performed. This is comparable to reading the back cover of a book, which gives insights into the story that will unfold, but without revealing the techniques used by the writer. Compared to prior work, this last quadrant might hold the most novel and promising applications, offering the possibility to \textbf{empower users} by drawing them out of their comfort zone and opening their mind to new situations. In PIF, we could, for example, implicitly measure a user's state to automatically shape a story and reinforce the ``sense of wonder'' that the author may wish to convey. This could help put back some magic in the world \cite{Rose2014}!

This taxonomy informed the design of PIF, so that it can be used to navigate between the quadrants and support a variety of applications: as text-based biofeedback, as a neuroadaptive tool to present content tailored to a reader's needs, to convey specific messages, or to enable people to put themselves in someone-else's shoes for the duration of a story. This last example was paramount in the development of PIF, as we believe that this type of application represents a tremendous potential in empowering writers and readers alike. As a first step towards this goal, we present the design and implementation of the PIF software tool.

\section{Design and Implementation}

\label{sec:implementation}

PIF was designed to propose applications along the four quadrants of the Taxonomy while providing freedom to content creators. The tool consists of four components (see Figure \ref{fig:archi}), each supporting one of the following design choices:

\begin{itemize} %

  \item To ease the authorship of interactive fiction, the engine is based on a markup language which enables story branching.

  \item To leverage on the flexibility of text and allow experienced content creators to incorporate other media, the software rendering the story is programmed with a widely used immersive platform.

  \item To access the reader's states and use it as input, a dedicated module processes the physiological signals.

  \item To alter the narrative and potentially affect the reader, an entity directs the story branching according to the writer's decisions. 

\end{itemize}
The communication between those four components relies on LSL \cite{siteLSL}, a protocol dedicated to the streaming and synchronization of physiological signals. LSL enables the various components of PIF to run on different machines within the same network to balance computational power.

\subsection{Branching the Story}

Stories in PIF are written in Ink, which handles both variables and conditional branching. Ink syntax is easy to learn and its various built-in mechanisms enable the creation of complex narratives. In Ink, readers progress in the narrative by explicitly selecting a course of action from a list when prompted. In PIF, manual choices can be replaced by automatic ones based on physiological signals, depending on ideas and feelings what the author want to convey or highlight. For example, the tone of a description can match the reader's emotion, or an event in the story can be triggered by their breathing rate. We added to the Ink syntax a tagging mechanism that let writers define contexts during which physiological data is recorded. For example, to record signals in a separate set of variables when part of a story occurs in a dungeon, keywords such as \verb!##DUNGEON_START! and \verb!##DUNGEON_STOP! can be added to delimit corresponding text passages. Thanks to a ``Director'' that updates all variables accordingly, Ink branching rules can then use this information to compare the reader's state between various contexts. For instance, the system can compare the reader's arousal between the said dungeon and a forest before deciding from where danger will ``unexpectedly'' arrive.

\subsection{Rendering the Story}

The rendering of the text is programmed in Unity \cite{siteUnity}. A virtual ``terminal'' (Figure \ref{fig:teaser}) is used as a canvas to display the stories and to give readers a spatial frame of reference. This asset can be incorporated within an existing Unity scene. Its appearance can be modified to better suit the needs of the author and can be incorporated in both 2D and 3D projects.
The terminal supports rich text formatting through HTML tags. Advanced users can change the text appearance by applying shaders. Besides added flexibility, this feature can be used to convey biofeedback with the text animated based on the user's physiological signals. PIF rendering can be deployed on multiple platforms, from a regular reading app on a tablet to an immersive virtual reality scene. In addition, the user's behavior in the scene can be recorded to gather information, such as a change in camera orientation. Also, LSL markers can be sent to help direct the story based on Unity Events. This allows for PIF to blend explicit actions taken within the rendering engine with the processed physiological data. %

\subsection{Physiological Signal Processing}

PIF collects physiological data that can be processed to assess the impact of text stimuli on the reader. Specific LSL markers are used to tag which story, branch, and page are currently being read. Using eye tracking, the system can detect the current word being read. 
For real-time analyses, the current implementation of PIF uses OpenViBE \cite{siteOpenvibe}, an open-source software with a graphical programming interface that provides various filtering techniques and classification algorithms. Thanks to the increasing number of devices supported by LSL and OpenViBE, the system can accommodate many sensors, medical or consumer grade, such as: heart-rate, breathing, electrodermal activity, or brain recordings. The output of the physiological signal processing pipeline is sent back through LSL and can then be used to select branches within the story.

\begin{figure}[t]
\includegraphics[width=1\hsize]{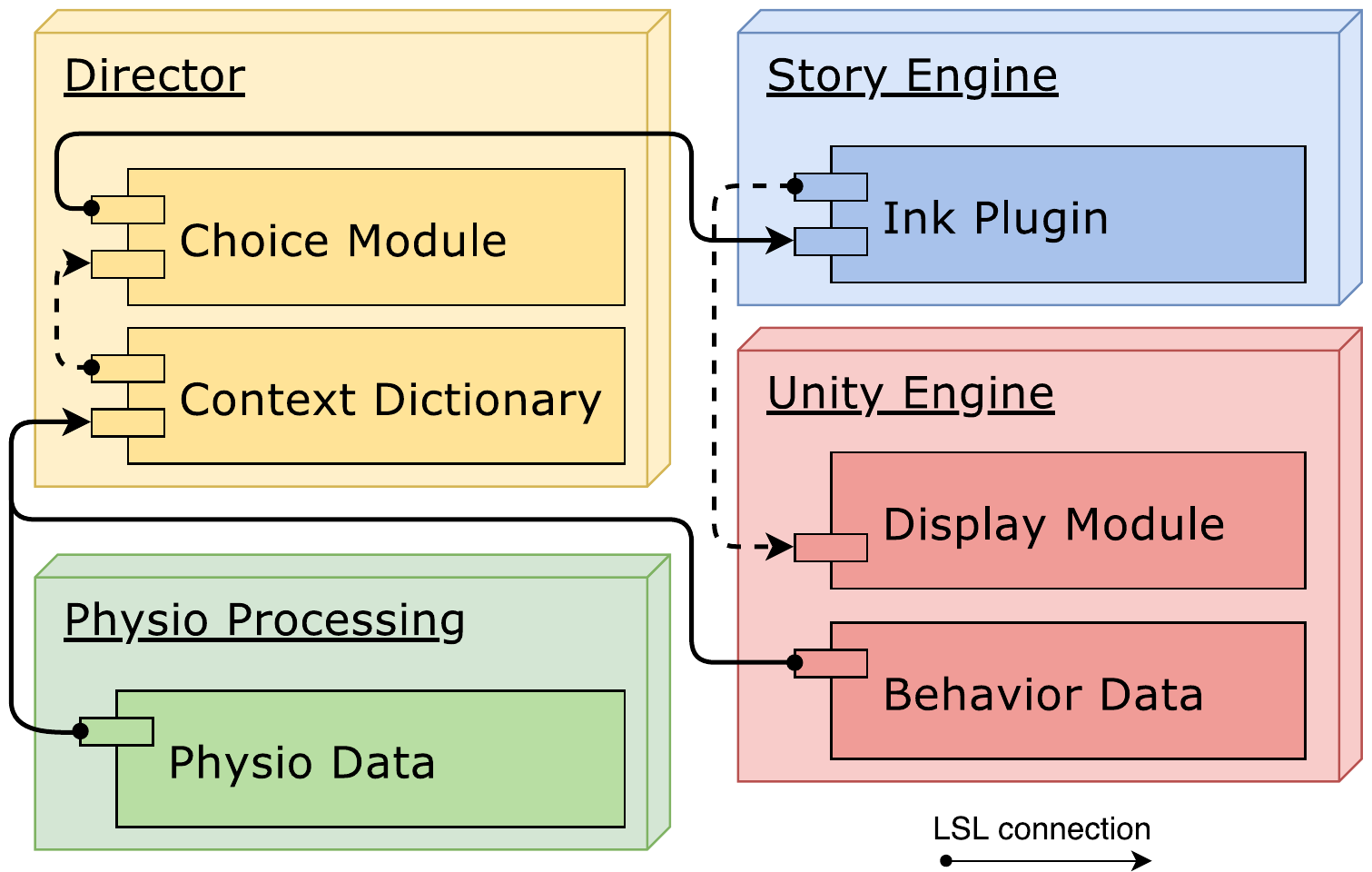}
\caption{Implementation architecture of PIF: a Unity module displays the text while a Director selects branches within a narrative written with Ink, using input from physiological or behavioral data. }\label{fig:archi}
\end{figure}

\subsection{Directing the Story}

Inspired by prior work, the ``Director'' (C\# project) links the physiological and Ink components. In game research, the Director is a high-level component monitoring a player's actions to orchestrate events and improve the game experience \cite{Yannakakis2013}. In educational research, intelligent tutoring systems demonstrate how rule-based systems improve content delivery to learners by adapting the difficulty of the exercises \cite{Murray1999}.
Here, as the story unfolds, the Director takes the processed physiological signals as input and updates the Ink variables in real-time with the data retrieved from LSL. If a section of the story is associated with a tag, the Director records and averages physiological data in a separate set of variables.

\subsection{Writing with PIF}

The following example provides insights about how the different PIF components work together. It helps better situate the role of each component and how authors can benefit from them. Let's consider a writer who would like to leverage the reader's reaction towards cats and dogs -- both animals with lovers and haters. The reader is equipped with a breathing sensor which provides the breathing rate, a marker of valence \cite{Boiten1994}, which is computed by OpenViBE. In Ink, tags are used to mark text involving either a cat or a dog. Each time one specie or the other is being mentioned, the Director is being notified through LSL markers. It then records the associated valence information to infer the reader's feelings towards each animal separately. Hence, upon reaching a branch, e.g., ``take care of an abandoned \textit{puppy}/\textit{kitten}'', it is possible to select the character's action that would induce the most, or the least, valence depending on the writer's intentions to please or challenge the reader. Such intentions are encoded using Ink syntax and the variables that are automatically updated by the Director.

We ran two user studies to determine how the system can be used to build physiologically driven interactive narratives. The first study investigates aspects of a story that can induce empathy and the second study assesses how a set of physiological features can be used to estimate the reader's state.

\section{Impact of Story Characteristics on Experience}

Prior work showed that a story can be perceived differently depending on the reader's own experience and demographics \cite{Cohen2001a, Hoeken2016}. This effect has been put in practice, for example by Robinson and Knobloch-Westerwick \cite{Robinson2017}, who investigated how 
readers perceived differently statements about sleep hygiene when characters were likable or not. In Communication, the effect or perceived similarity with characters has been more broadly described. Moyer-Gusé and Nabi \cite{Moyer-Guse2010} studied how entertainment TV programs could induce persuasion about health and social issues when the characters and the viewers are more alike. This proximity between a reader (or a viewer) and a narrative can greatly impact how it is interpreted since identification is a natural mechanism to expand the boundaries of self \cite{Slater2014}.

Going toward the \emph{Empowering} quadrant of the Taxonomy, we envision PIF as a way to immerse readers in foreign situations and empathize with characters. For instance, one might want to relate to the story of a person during their first day at work. The main protagonist possesses several features, such as gender, type of work, or where they grew up (e.g., rural or urban area). Each combination will result in a different version of the story, that may move readers differently. To raise awareness about gender inequalities, male engineers who grew up in the countryside could, for example, read a story about a woman who moved to the city for an internship in the tech industry, and who is facing difficulties in a male-dominated workplace.

In contrast with prior work, using implicit input for IF requires finding characteristics that can also be measured in real-time, such as those related to emotions. To determine how the background of a reader can impact their reading experience, we conducted an online study which we present below.

\subsection{Online User Study}
To investigate story characteristics that can engage readers, we created two short stories with elements intended to elicit strong responses. Each story contains two varying elements, thus, there are 2 $\times$ 2 versions of each story. Our hypothesis is that different versions of a story will impact readers differently. We are interested in finding characteristics that can significantly alter a reader's perceived similarity and engagement, once matched with their background and preferences. The results will serve as guidelines for real-time text adaptations.

In one story, the protagonist intends to commit a robbery to escape an abusive father, and to do so they break in the house of an elderly person who just lost their only child. When the main protagonist faces spider webs, they can react in two ways: be frightened at the sight of the arachnid, or continue nonchalantly. Depending on reader's aversion toward spiders, one of the alternatives would match their own reaction. Besides the reaction to the spider, the settings of the first plot can either be an urban or a rural environment. In a second story, a person has their bike stolen while they are seated at a coffee shop, preparing for an important presentation. The protagonist can either be male or female, and the subject of the presentation is related to either computer science or psychology. %

The choice of stories characteristics were based on literature \cite{Hoeken2016} and on the fact that they could be measured through physiology (e.g., increased arousal caused by the fear of spiders). The number of stories and characteristics were limited so that reading time would be no more than several minutes. The study was designed to last for a total 30 minutes to avoid fatigue.
To measure the readers' reactions to the depicted events, two questionnaires were administrated immediately after each story: the narrative engagement questionnaire \cite{Busselle2009} that assesses how well people are immersed in a narrative (scale 0 to 72); and a measure of perceived similarity (``perceived homophily'') \cite{McCroskey1975} where people rate how much another person is similar to them (scale 7 to 84), which was used as a proxy for empathy. 
To alleviate the variability inherent to online studies and ensure that readers understood the stories, an English proficiency test, Lextale \cite{Lemhofer2012}, was administered. Additionally, to control the readers' attention, each story was followed by two comprehension questions. Finally, a questionnaire assessed how much the reader's background matched the narratives, including demographics related to where they grew up, their field of expertise, their fear of spiders, if they ever lost someone close to them, and if an object was ever stolen from them.

\subsection{Protocol}

The survey was distributed via the Amazon Mechanical Turk platform which allows workers over 18 y.o. to complete online tasks for pay. Workers were sampled with an excellent performance history, HIT approval rate $\geq$ 97, an approved number of HITs $\geq$ 100, and restricted to geographical areas where English is the primary language. %
Participants read the 2 stories in a randomized order, each story randomly selected among 4 possible versions. After each story, participants filled in the narrative engagement and similarity questionnaires, as well as the comprehension questions. The survey ended with the demographics questionnaire and Lextale test.

Participants were compensated US\$3.5 for a survey that took in average 30 minutes to complete. A total of 168 approved volunteers were sampled for this study. Our inclusion criteria were: a score on the Lextale test $\geq$ 80\% (C1/C2 proficient speakers on the Common European Framework \cite{Lemhofer2012}), correct answers to the 4 comprehension questions, and a reading time of at least 30 seconds for each story. Twenty-eight (28) participants did not meet those criteria and as a result 140 participants (56f, 84m) were included in the data analysis.

\subsection{Data Analysis}

To investigate readers' perception of the narratives, we computed whether or not the characteristics of the stories were similar to the information participants provided about their background. For the first story, we defined three characteristics: location (rural or urban), bereavement, and aversion to spiders. For the second story the three characteristics were: gender, field of expertise, and having had a personal object stolen. %
We used stepwise regressions with Akaike Information Criterion (AIC) optimization to model the relationships between those characteristics and the measurements of narrative and perceived similarity. As compared to other criteria, AIC is best suited to find a model that is robust to new data \cite{Kuha2004}. A stepwise regression works by incorporating factors one by one, using AIC to estimate and compare the models. Only factors that increase the model's robustness are kept.

\begin{table}
  \caption{Parameters of the regression models predicting perceived similarity and narrative engagement for the 2 stories, based on the match between story characteristics and readers' backgrounds. Empty cells represent coefficients not kept in the corresponding regression models.}
  \label{tab:mturk}

\begin{tabular}{ccccc}
    \toprule
     Dimension & Intercept & Fear & Location & Loss  \\
    \midrule
    Similarity & 34.4 & 3.5 & & \\
    Engagement & 44.9 & & 3.5 & \\
  \bottomrule
\end{tabular}

\vspace{0.2cm}

\begin{tabular}{ccccc}
    \toprule
     Dimension & Intercept & Gender & Expertise & Stolen \\
    \midrule
    Similarity & 49.0 & &  & 4.3 \\
    Engagement & 44.1 & &  & 4.4 \\
  \bottomrule
\end{tabular}

\end{table}

\subsection{Results and Discussion}

The results of the different regressions are presented in Table \ref{tab:mturk}. We found that narrative engagement increases when the location depicted in the story matches the environment readers grew up in. Narrative engagement, as well as the perceived similarity of a story character, increase when there is a negative feeling associated with an object being stolen. Finally, perceived similarity also increases if the character reacts as the reader would when facing a spider -- i.e., the character stays calm if the reader is not afraid of spiders or panics otherwise. Story characteristics such as the presence of spiders or stolen objects are interesting because physiological sensors can be used to measure the emotions related to them -- here, arousal and valence. Hence, adapting the story depending on a reader's state could be a strong tool to elicit engagement or empathy toward a character. We suggest that a writer could decide to detect how the reader reacts to various textual stimuli, and reflect the reader's reactions through the main character. Such adaptation could be further reinforced with other writing techniques, such as shifting the point of view of the story to present others' perspectives \cite{Graaf2012}.

In this study, we investigated how story characteristics can foster narrative engagement and increase the perceived similarity with a story characters when they are matched to the reader's background. To replace the questionnaires we administered and assess the feasibility of real-time adaptation, the next section investigates the possibility to measure the reader's state through physiological signals alone.

\section{Understanding readers reactions}

The previous section determined that specific story characteristics can impact the reader's experience. To use this in real-time for interactive fiction, we first validate how physiological data can be used as implicit input. This section presents a user study that was conducted to assess how PIF can be used to measure a reader's physiological reactions to a story. 

Among the physiological sensors commonly used in HCI, we chose to monitor breathing, ElectroDermal Activity (EDA), pupil dilatation, and eye movement. Brain recordings were not considered as they suffer from high inter-subject variability and require a long setup from experts, as well as gel-based electrodes, to acquire a good quality signal (see \cite{Nijboer2015}). Similarly, heart rate measurements would have required to setup electrodes to get reliable measures, with little benefits compared to the other sensors at our disposal. Current wearable devices, such as smart watches, have yet to demonstrate their ability to consistently account for heart rate variability to supplant breathing or EDA over short recording sessions. 

The combination of breathing, EDA, and eye tracking was the best trade-off between user comfort, reliability, and users' states measurement. Breathing features are an indicator of valence \cite{Boiten1994}, EDA is an indicator of the autonomous nervous system activity \cite{Cowley2016}, and gaze fixations and saccades are indicators of reading difficulty \cite{Rayner2009}. Using these sensors together to classify reactions to stories, our hypothesis is that we can differentiate between: \emph{positive} and \emph{negative valence}; \emph{low} and \emph{high arousal}; and texts that are either \emph{simple} or \emph{difficult}.

We designed a 3 $\times$ 2 within-subject experimental study where participants read 3 pairs of texts that were likely to trigger strong reactions. We created a signal processing pipeline to classify participants' reactions during the reading session. 

\subsection{Experimental Setup}

In the experiment, we compared six texts divided in the following three pairs of stories:

\begin{itemize}

\item A \emph{happy} story: a son of a wealthy family doing charity work and realizing the value of life; and a \emph{sad} one: a child waking up from a nightmare only to meet his parents' murderer.

\item A \emph{boring} story: a bunny wandering in a clover field; and an \emph{exciting} one: a driver trying to avoid a fine by deceiving a police officer.

\item A \emph{complicated} story: a scientific description of a spectrometer; and a \emph{simple} one: a guide on how to make coffee.

\end{itemize}

Aiming at a responsive system that could adapt within moments to a change in the reader's states, we opted for short stories. To limit confounding factors, the stories were of the same length (1469 to 1522 characters), and all were displayed across 4 pages on the PIF terminal screen (Figure \ref{fig:teaser}) using the same font. An additional neutral ``dummy story'' was written to be used as tutorial at the beginning of the experiment. 

We used materials presented in the \nameref{sec:implementation} section to display stories and acquire physiological data. The stories were displayed within a 3D scene through a high resolution (2560x1440 pixels) head-mounted display with embedded eye tracking (FOVE 0 \cite{siteFove}). This setup was chosen to reduce variability across participants by controlling the user's field of view. The terminal was positioned 1m in front of the viewer in the virtual space with its display representing a surface of 0.8m x 0.6m, mimicking the presence of a screen. Using the FOVE, we gathered head rotations, pupil dilatation, as well as the convergence point of both eyes onto the plane where the text was displayed. The eye tracking data was acquired at 70Hz, corresponding to the refresh rate of the FOVE, and the data was streamed from Unity through LSL (see Figure \ref{fig:archi}).

Breathing was measured using a g.tec \cite{siteGtec} g.RESPsensor -- a belt which comprises a piezo-electric sensor reacting to chest inflation -- and EDA with a g.GSRsensor. Both sensors were connected to a g.USBamp amplifier. The signals were acquired at 512Hz on a dedicated machine with OpenViBE acquisition server and streamed through LSL on the local Ethernet network. All signals and markers were recorded for offline analyses using LabRecorder.

\subsection{Protocol}

We recruited fourteen participants (11m, 3f) between 21 and 31 years old (mean=25.0, SD=3.1) who were compensated the equivalent of US\$8 in local currency for their time. Participants were fluent English speakers and none wore glasses.

Participants were welcomed in the experimental space. After signing a consent form, they were asked to fill in a demographic questionnaire. They were then equipped with the breathing belt and the EDA electrodes placed on the index and medium fingers of the non-dominant hand and secured with skin tape. Participants were seated on a comfortable chair and asked to stay still throughout the recordings. After verifying that the physiological sensors were adequately connected, participants placed the headset on their head and proceeded to a short ($\approx$ 10 seconds) eye-tracking calibration.

To ensure that each participant understood the procedure and that the equipment was properly set up, they were guided through the dummy story before the start of the recording.
While the participant's non-dominant hand was resting on a flat surface to avoid EDA artifacts, their dominant hand was placed over a keyboard to advance between pages at their own pace. A 2-second time-out prevented them to accidentally double-press on the keyboard.

The six stories were displayed in a random order and a blank page was announcing the beginning of a new story. Users kept the head-mounted display on between the stories to maintain the calibration and ensure reliable eye tracking. Participants were notified when they reached the half-way point of the study. An attention questionnaire was administered at the end of the study to ensure that participants were attentive to each story. It consisted in a comprehension question with four choices per story. The reading session lasted approximately 10 minutes and the total experiment time was under 30 minutes.

\subsection{Signal Processing}

\begin{figure}
\includegraphics[width=1\hsize]{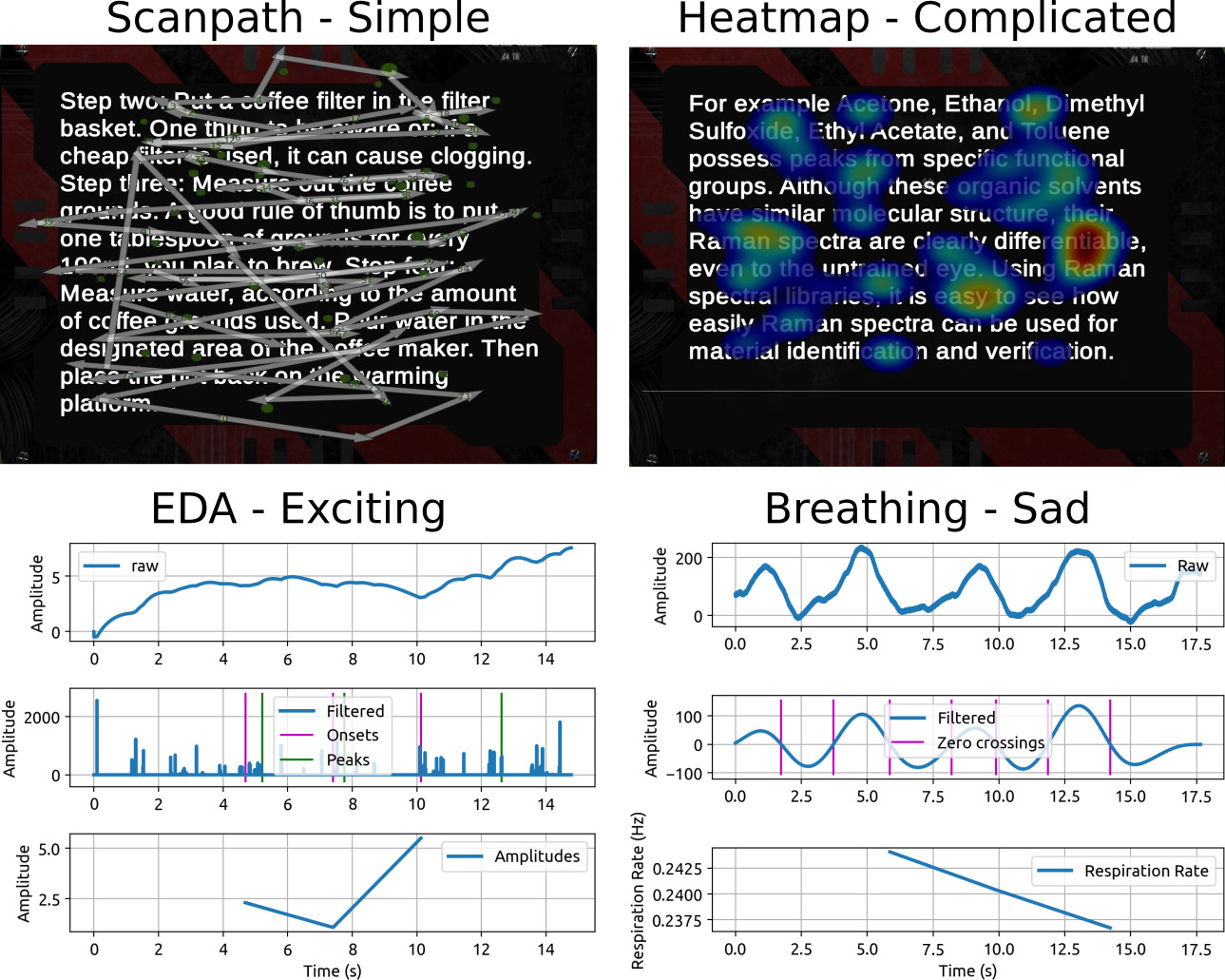}
\caption{Example of physiological data extracted while reading. Top left: scan path presenting eye saccades from the page of the \emph{simple} text; Top right: heatmap of eye fixations during the \emph{complicated} text; Bottom right: increased EDA activity during the \emph{exciting story}; Bottom left: reduced breathing rate during the \emph{sad} story.}\label{fig:features}
\end{figure}

We recorded one set of physiological recordings per participant (see Figure \ref{fig:features}) and per story (70 seconds in average). The signals and extracted features are described below.

\subsubsection{Eye Gaze Data}

We processed gaze data using Python PyGazeAnalyser \cite{Dalmaijer2014}. All points that were off by more than 10\% from the location of the virtual terminal were discarded. An \textbf{eye blink} was counted when there the was no point on-screen for a duration between 50ms and 500ms \cite{Caffier2003}. We counted the \textbf{number of eye blinks} and \textbf{average duration}. We defined as \textbf{mind wandering} the total amount of time the reader was not looking at the virtual display during a story. It was computed by summing events during which there were no points on-screen for a duration greater than 500ms (upper limit of a blink). We computed the \textbf{number and average length of fixations and saccades}, as well as the \textbf{average angle of the saccades}.
To better reflect how fluently readers scanned the text, we also extracted \textbf{split saccades} features using an upper limit of 350ms \cite{Baloh1975}.

\subsubsection{Pupil Dilation}

Pupil dilation is an indicator of workload and emotions \cite{Cowley2016,Partala2003}. In the Fove SDK, pupil dilation is represented by a relative value, with "1" corresponding to a baseline computed during the calibration phase. We extracted the mean value and standard deviation as pupil dilation features.

\subsubsection{Head Movements}

From head rotations' quaternions, we extracted the total travel distance, computed as the sum of the geodesic distance between each sample, as well as the average speed -- travel distance divided by recording length.

\subsubsection{Electrodermal Activity}
Within the phasic component of EDA, ``peaks'' can appear in the signals a few seconds after the presence of a stimulus. The total number of peaks and their average amplitude were extracted using Biosppy. The average SudoMotor Nerve Activity (SMNA), a robust discriminator of the autonomous nervous system, was computed using the cvxEDA library \cite{Greco2016}.

\subsubsection{Breathing}

We used the Python libraries Neurokit \cite{siteNeurokit} and Biosppy \cite{siteBiosppy} to extract breathing features. We applied a band-pass filter between 0.1Hz and 0.35Hz before computing the breathing rate, as well as the breathing RMSSD (Root Mean Sum of the Squared Differences), an indicator of variability. Since the piezo-electric component of our breathing sensor reacts to \emph{changes} in thoracic circumference (i.e., the signal would decrease while holding breath), we extracted a second set of features after performing a cumulative integral. 

\subsubsection{Classification}

We used Weka \cite{siteWeka} to perform binary classifications within each pair of stories, i.e., \emph{happy} vs \emph{sad}, \emph{boring} vs \emph{exciting}, and \emph{complicated} vs \emph{simple}, using the features described above. Yannakakis \& Togelius \cite{Yannakakis2017} described how we tend to rate emotions by comparing one with another, rather than assigning absolute values. Hence, to reflect how we represent our inner states, features were normalized per subject by converting numerical values to ranks, drastically improving the efficiency of our pipeline.

\begin{figure}[t]
\includegraphics[width=1\hsize]{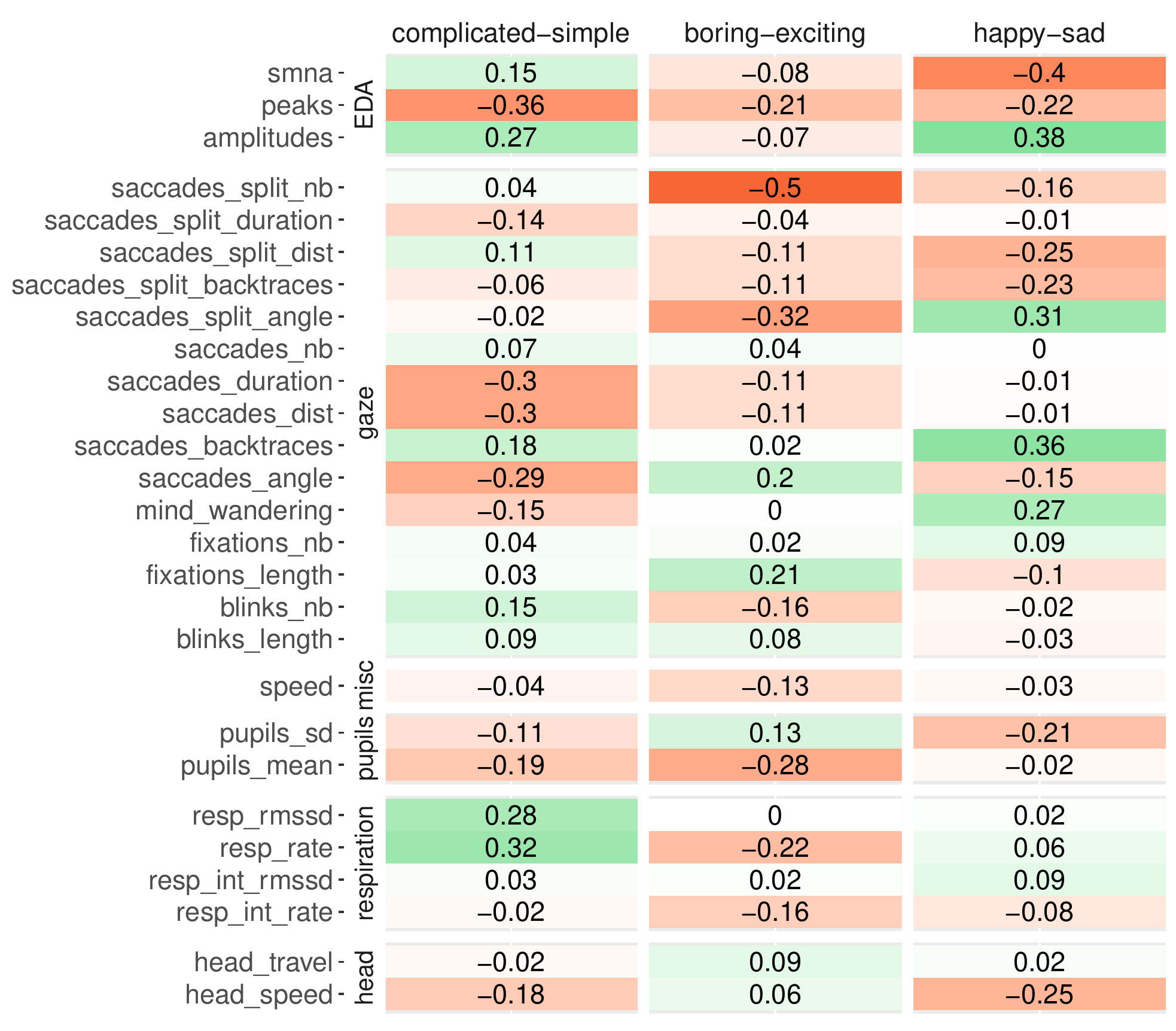}
\caption{Average weight of each feature after training the three classifiers. The values are normalized between 1 (stronger association with the first story specified in the column) and -1 (stronger association with the second story).}\label{fig:weights}
\end{figure}

To reduce complexity and improve classification, we performed a Principal Component Analysis (PCA) on the 27 features (Figure \ref{fig:weights}), retaining 95\% variance. 

We then used Fisher's Linear Discriminant Analysis (LDA) to classify data. To simulate an online scenario where the physiological signals of a user would be classified using training data gathered with past participants (i.e., transfer learning), classification accuracies were computed using ``leave one subject out'' cross-validation. To do so, classifiers were tested on each one of the fourteen participants while being trained on the remaining thirteen ones.

\subsection{Results and Discussion}

In the attention questionnaire, participants obtained on average 90\% correct answers (SD=11 percent point) with:
\begin{itemize}
    \item \emph{complicated}: 69\% (SD=6pp) vs. \emph{simple}: 100\% 
    \item \emph{boring}: 77\% (SD=6pp) vs. \emph{exciting}: 100\%
    \item \emph{happy}: 100\% vs. \emph{sad}: 92\% (SD=4pp)
\end{itemize}

These numbers suggest that participants were attentive to the events taking place in the narratives, except for the \emph{complicated} and \emph{boring} narratives where they could not always recall the details of the stories. The classification accuracy are reported in Table \ref{tab:res}. 

Figure \ref{fig:weights} presents the weight each feature was given after the training of the 3 binary classifiers, averaged and normalized across participants. Weights range from 1 to -1, where a value close to 0 means that the feature does not discriminate well between the two classes. Positive and negative values mean that the feature is more predictive of either one class or the other. For example, the negative weights attributed to the EDA features in the \emph{Arousal} classifier (i.e., boring-exciting) confirm that increased EDA is associated with more exciting stories. 

Considering the absolute weights, we found that the features extracted from pupils and gaze were used the most by the \emph{Arousal} classifier. We also found that EDA and breathing were the two most important features for the \emph{Difficulty} classifier (i.e., complicated-simple). Finally, EDA and head movements were most prominent for the \emph{Valence} classifier (i.e., happy-sad).

We used the PIF engine to gather physiological data and compare reactions to textual stories on three scales: Arousal, reading Difficulty, and Valence. We built classifiers to determine a reader's reaction to a story after approximately 3 minutes of reading. Among our three classifiers, Arousal, comparing \emph{boring} vs \emph{exciting} stories yielded the highest performance with 92.9\% classification accuracy. The reading Difficulty classifier comparing \emph{complicated} vs \emph{simple} texts presented good accuracy (78.6\%). The Valence classifier comparing \emph{happy} vs \emph{sad} tones did not perform as well, and was only marginally better than chance with 64.3\% classification accuracy. 

The weights attributed by the classifiers to input features can be used by writers or researchers to determine sensors of interest to investigate a particular construct, such as using pupil dilatation for arousal. Our results exposed that EDA was more prevalent in the \emph{Valence} than in the \emph{Arousal} classifier, which contrasts with the literature \cite{Cowley2016}. This might be due to a discrepancy between the emotions that we intended to elicit (as writers) and the actual perception of the study participants (as readers). 

In the future, we suggest building a database of labelled short texts %
that could be used with PIF to train the classifiers. The collection of background information on participants would help better understand how narratives are intertwined with users' states, especially since there is no direct and one-way causal link between reading a text and a change in physiology. Additional data can also be used to investigate how users would react over time when the interactive narrative spans over several days or weeks of reading.
While data was processed offline to validate our approach, the python libraries used for feature extraction, as well as the classification algorithms, can be integrated directly within OpenViBE, facilitating the transition toward online processing. In this experiment, readers were exposed to short stories with a total reading time below 10 minutes and each trial lasting 70 seconds in average. Should PIF be used to read an interactive novel, which reading spans over hours, we anticipate the robustness of the system to increase by an order of magnitude. For more precise measures, we envision replacing the binary classifiers with algorithms allowing for probabilistic outputs, such as Bayesian classifiers.

\begin{table}
\centering
  \caption{Classification Accuracy Between Pairs of Textual Stories}
  \label{tab:res}
  \begin{tabular}{cc}
    \toprule
    Measured Construct (Stories) & Accuracy \\
    \midrule
    Arousal (Boring \emph{vs} Exciting) & 92.9\% \\
    Difficulty (Complicated \emph{vs} Simple) & 78.6\% \\
    Valence (Happy \emph{vs} Sad) & 64.3\% \\
  \bottomrule
\end{tabular}
\end{table}

Overall, these results demonstrate the feasibility of using physiological signals as input to assess the reader's experience. These findings can be used to explore the ``Empowering users'' quadrant of the taxonomy, using physiology as implicit input with PIF to perform automatic adaptation of an entire story.

\section{Discussion}

We built upon the concept of interactive fiction and proposed the use of physiological signals to drive a narrative. We demonstrated the feasibility to automatically determine a reader's inner states by monitoring their physiology. We proved that within a single minute of reading, the reader's level of arousal and the reading difficulty can be inferred using physiological signals. We advance that additional applications can be achieved depending on whether readers explicitly affect their physiology, and whether they are aware that the system may be affecting them. We further propose that such measurements can be used in broader applications areas.

We defined a taxonomy of physiologically driven interactive systems which considers physiology as both input and output along two dimensions: awareness and explicitness, with regards to the user. This taxonomy will support future research using physiological signals in interactive systems. In its current state, it helped us bring to light important ethical concerns in how physiological signals can be used. For instance, how should we handle a system that would covertly manipulate a person's state? History tells us that storytelling has been used over the ages as a tool to communicate ideas, concepts, and lessons. Yet, it has also been used for propaganda. Beyond ethics, we foresee that awareness of how an interactive story could affect the reader's state is crucial for its acceptability. %
Once people are aware of the possible uses and missuses, the concept of physiologically driven storytelling opens up novel opportunities to positively affect readers. The PIF software tool enables the design of novel applications, such as a ``narrative biofeedback'' to support people in regulating their anxiety, or the development of an adaptive system to improve a person's comfort by adapting the readability of a text in real-time. Compared to traditional interactive fictions, using physiology as implicit input for adaptation could prevent breaks of presence, which occur when one shifts their attention from an immersive environment to their physical surroundings \cite{Slater2009}.

PIF, open-source: %
\url{https://pif-engine.github.io/}, can be used to explore all quadrants of the Taxonomy and further investigate fundamental research questions.
The dissemination of robust sensors, such as the detection of muscle activity and facial expression in head-mounted displays, %
or wearables with medical-grade heart-rate measurements, will likely attract new users. We believe these different resources can help bring a new community together around the use of physiological signals in interactive systems. 

\section{Future Work}
With, on the one hand, a set of features that elicit engagement (online study), and, on the other hand, a set of features that can be recorded through physiology (laboratory study), the results gathered in this research work are oriented toward the \textit{Empowering} quadrant of the Taxonomy. Our next step is to combine the results from the online and laboratory studies to investigate the extent to which physiology can be used to modify a person's state. It would, for example, be worth investigating whether identification toward characters leads to increased empathy and whether such system could be used to increase empathy in support of conflict resolution.  
In addition, based on our results, a promising research direction would be to work towards increasing the resemblance between the reader and the main character, such as by detecting aversion to some stimuli (e.g., spiders), or detecting the reader's current mood and altering the narration accordingly. %
Through increased identification, we hope to leverage compassion.
In the future, PIF could be used to turn stories into persuasive technology, such as for well-being as envisioned in the science fiction literature, as in Ender's Game \cite{Card1985} and The Diamond Age \cite{Stephenson1995}. %
We further imagine PIF as a tool to assist journalists and storytellers when discussing societal issues such as racism and gender discrimination, helping readers empathize with foreign situations. This work is a first step towards building empathizing authoring tools.

Additional sensors and signals can be incorporated in PIF. Since OpenViBE was designed to process brain recordings, this signal would be easy to include in the system, even though it was not considered in this first iteration due to their inherent complexity. In the future, BCI can give valuable insights about a user's states, as demonstrated in past work aimed at assessing user experience \cite{Frey2016}. In addition, we intend to further evaluate the PIF software tool by conducting usability workshops with authors as well as focus groups with readers.

\section{Conclusion}
In this paper, we put forth the concept of physiologically driven storytelling. We then described a taxonomy of use of physiological signals in interactive systems and showed that depending on whether physiology is used as input or output, and on the user's awareness, four types of use-cases can be created: \emph{biofeedback}, \emph{neuroadaptive systems}, \emph{deceiving users}, and \emph{empowering users}.
We then presented the design, implementation, and evaluation of PIF, a software tool for Physiological Interactive Fiction. This tool allows to synchronize a narrative with the reader's physiological data, so that the story can be adapted to the readers' inner states. PIF was designed to support the exploration of the four taxonomy quadrants. We proposed that most novel applications relate to users' empowerment when the system helps readers empathize with foreign situations or characters. 

We conducted two user studies, one to uncover story characteristics that can elicit engagement and increase perceived similarity, and a second to assess what states can be measured when reading a story and which physiological features are the most useful to do so. We found that constructs linked to cognitive processes %
resulted in higher classification accuracy than constructs related to affect. %
The PIF software tool has been made freely available for future research and for writers to explore novel ways to engage readers. 
This work opens the space to future physiologically driven applications and systems within broader application areas, such as behavior change and the promotion of positive emotions.

\section{Acknowledgments}
The authors thank Doron Friedman and Jonathan Giron for hosting the experiment in their facilities, Jelena Mladenovi\'{c} for her help on the visual material, as well as Danielle Goldberg and Amir Lorch for editorial feedback.

\pagebreak

\balance{}

\bibliographystyle{SIGCHI-Reference-Format}
\bibliography{biblio}

\end{document}